ARMA 14-7442

# Fracturing tests on reservoir rocks: Analysis of AE events and radial strain evolution

Pradhan, S.
*SINTEF Petroleum Research, Trondheim, Norway*
Stroisz, A.M., Fjær, E., Stenebråten, J., Lund, H. K., Sønstebø, E. F.
*SINTEF Petroleum Research, Trondheim, Norway*
Roy, S.
*The Institute of Mathematical Sciences, Chennai, India*

**ABSTRACT:** Fracturing in reservoir rocks is an important issue for the petroleum industry - as productivity can be enhanced by a controlled fracturing operation. Fracturing also has a big impact on $CO_2$ storage, geothermal installation and gas production at and from the reservoir rocks. Therefore, understanding the fracturing behavior of different types of reservoir rocks is a basic need for planning field operations towards these activities. In our study, the fracturing of rock sample is monitored by Acoustic Emission (AE) and post-experiment Computer Tomography (CT) scans. The fracturing experiments have been performed on hollow cylinder cores of different rocks - sandstones and chalks. Our analyses show that the amplitudes and energies of acoustic events clearly indicate initiation and propagation of the main fractures. The amplitudes of AE events follow an exponential distribution while the energies follow a power law distribution. Time-evolution of the radial strain measured in the fracturing-test will later be compared to model predictions of fracture size.

## 1. INTRODUCTION

How one can fracture reservoir rocks efficiently without damaging the well or environment - is a big challenge to the petroleum industry. This problem is also linked to the implementation of underground $CO_2$ storage and geothermal energy production scenarios. The fracture initiation mechanism and propagation dynamics [1,2] in porous rocks need to be analysed and understood well for solving the problem and answering the calls – Where does fracture go in reservoir rocks? How does the fracture plane look like? How fast does the fracture move?

During fluid injection in reservoir rocks – mainly hydraulic fracturing occurs. Usually, a sudden increase in fluid pressure generates a hydraulic fracture, but sometimes stress drop also plays a key role. In a porous reservoir, fluid pressure can rise due to heating, gas generation, mineralogical changes, communication with another high pressure zone, or due to human activities associated with oil and gas exploration [1,2].

So far, modelling of fracture initiation and growth [1-4] has not been very successful as it is often based on linear elastic fracture mechanics, with resulting predictions that fail to reproduce reality. In this work we study fracturing in reservoir rocks through lab experiments. First we do fracturing test on hollow cylinder core samples under high injection pressure with AE monitoring system -that can locate the cracking events responsible for the fracturing process. AE data are recorded during the entire test until the main fracture opens up. Statistics of AE events – in terms of amplitude distribution and energy distribution -have been analysed for all the rock types. We also record radial strain of the rock sample during the test in order to compare the time evolution of radial strain with the predicted values from our model.

## 2. FRACTURING TEST

In this study, the fracturing of different rock types was obtained by injecting pressurized oil into a rubber tube fitted in the center of a hollow cylinder rock core. The tube prevents fluid to migrate into the sample during the test. The borehole pressure was enhanced gradually, upon 0.3 mm displacement of pump piston between each step, until failure occurs. Constant oil confinement of 5 MPa was exerted on an impermeable sleeve during the entire test. This tightens the sleeve around the sample adjusting the chain for radial strain measurements and improving the pinducers – sample contacts. A symmetrically distributed push-in type inserts were used to fix the position of nine to twelve pinducers at the circumference of the samples, at four levels along the length (see Fig. 1).

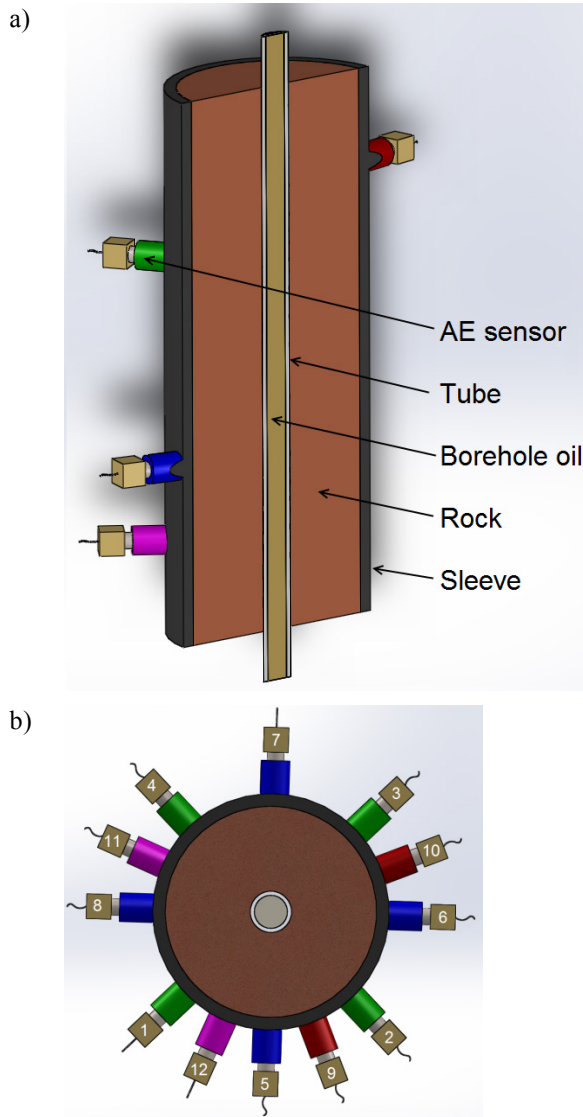

Fig. 1. Location of AE sensors in the setup shown in a longitudinal (a) and cross (b) section view. Sensors 1-4 (green) and sensors 5-8 (blue) are positioned at 23 mm from the center to the top and bottom, respectively; sensors 9-10 (red) and sensors 11-12 (pink) are at about 2/3 from the upper and lower edge of sample, respectively.

The acoustic emission (AE) activity, detected by the sensors, was registered with a multi-channel Vallen system supplemented with 34 dB preamplifiers. A detail description of the setup and test procedure has been given in [5].

Acoustic emissions detected during fracturing tests are elastic waves produced by sudden internal stress redistributions caused by changes in the rock's body. Such structural changes concern mainly crack opening and growth, dislocation movement, etc. The maximum AE activity is founded in the close vicinity of the peak stress at which the global fracture occurs (see Fig. 2).

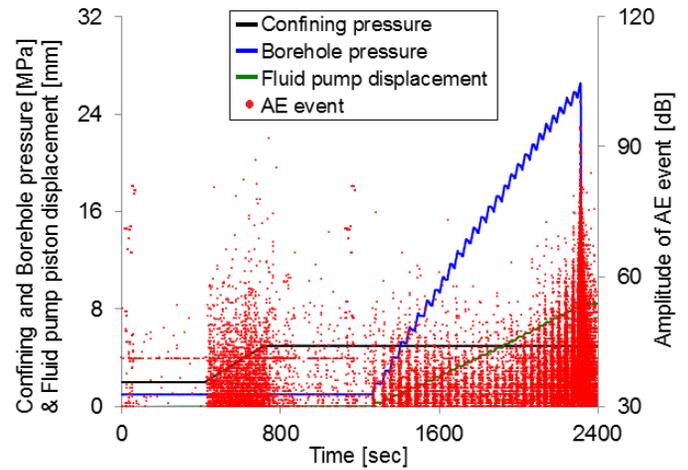

Fig. 2. Stress path and AE activity during fracturing test. Confining pressure (black), borehole pressure (blue), piston displacement of the fluid pump (green) and AE events (red points). Data refers to a test on Saltwash North sandstone.

Types of rocks tested in this study, with mineralogy and selected properties, are listed below:

*Berea sandstone* - composed of 80% quartz, 12% feldspar and rock fragments, and about 8% clay. The porosity is about 19% and density 2.2 g/cm$^3$. Young's modulus is ~ 13.5 GPa, unconfined strength ~ 42 MPa, indirect tensile strength ~ 4.7 MPa, and P-wave velocity ~ 2090 m/s.

*Castlegate sandstone* – composed of 70% quartz, 30% feldspar and rock fragments, and low clay content (often assumed as clay-free). The porosity is about 28% and density 1.9 g/cm$^3$. Young's modulus is ~ 3.4 GPa, unconfined strength ~ 20.4 MPa, indirect tensile strength ~ 0.9 MPa, p-wave velocity ~ 1830 m/s.

*Red Wildmoor sandstone* – composed of 42% quartz, 47% feldspar, 2% other rock fragments, 9% clay. The porosity is about 27% and density 1.9 g/cm$^3$. Young's modulus is ~ 3.4 GPa, unconfined strength ~ 19.3 MPa, indirect tensile strength ~ 0.9 MPa, p-wave velocity ~ 1590 m/s.

*Saltwash South sandstone* – composed of 84% quartz, 5% feldspar and rock fragments, and about 11% clay. The porosity is about 30% and density 1.8 g/cm$^3$. Young's modulus is ~ 0.3 GPa, unconfined strength ~ 1.9 MPa, indirect tensile strength ~ 0.2 MPa, p-wave velocity ~ 980 m/s.

*Saltwash North sandstone* – composed of 86% quartz, 9% feldspar and rock fragments, and about 5% clay. The porosity is about 21% and density 2.1 g/cm$^3$. Young's modulus ~ 3.0 GPa, unconfined strength ~ 20.8 MPa, indirect tensile strength ~ 1.7 MPa, p-wave velocity ~ 1300 m/s.

*Mons chalk* – composed of 99% calcite ($CaCO_3$), 1% quartz and pyrite inclusions. The porosity is about 44% and density 1.5 g/cm$^3$. Young's modulus is ~ 4.9 GPa, unconfined strength ~ 13.3 MPa, indirect tensile strength ~ 1.7 MPa, p-wave velocity ~ 2140 m/s.

*Lixhe chalk* – composed of 99% carbonate, 1% silica and clinoptilolite. The porosity is about 42% and density 1.5 g/cm$^3$. Young's modulus is ~ 4.7 GPa, unconfined strength ~ 10.0 MPa, indirect tensile strength ~ 1.2 MPa, p-wave velocity ~ 2320 m/s.

The samples were prepared as hollow cylinder plugs of 51 mm outer diameter, 10.5 mm inner diameter, and 135 mm length, approximately. All were tested dry, after 48 hours drying at 120$^O$C. At least two samples for each rock type have been examined – sixteen samples in total.

## 3. STATISTICAL ANALYSIS OF AE DATA

Analysis of Acoustic emission (AE) signals during the fracturing tests can help understanding the details of rock-micro-fracturing and fracture propagation. AE studies utilize hypocenter mapping, event statistics and focal mechanism to investigate crack formation and propagation, damage precursors and failure modes of material/rock samples under compression or external loading. AE studies [6,7] for compression test on dry and wet sandstone reveal that micro-fracturing is actually controlled by the amount and distribution of weak minerals. A similar test on granite [8] has identified a zone of distributed micro cracks (process zone) around the tip of propagating fractures and the recorded data shows that the density of micro cracks and amount of AE increase while approaching the main fracture. Another AE study on sandstone under hydrostatic and triaxial loading conditions [9] confirms the formation of compaction bands during the fracture process. In case of fracturing in composite materials under external stress AE bursts follow universal power law statistics – that has been observed in numerical models [10] and explained/confirmed by theoretical calculations [11,12].

During the entire fracturing test we recorded AE events (Fig. 2). In all the cases, the event rate increases as we approach the final fracturing point. This feature is quite common in all the fracture models [3,4]. We have studied the statistics of AE amplitudes and energies recorded at different AE receiver channels (CH) of the Vallen AE monitoring system. Two examples (one for

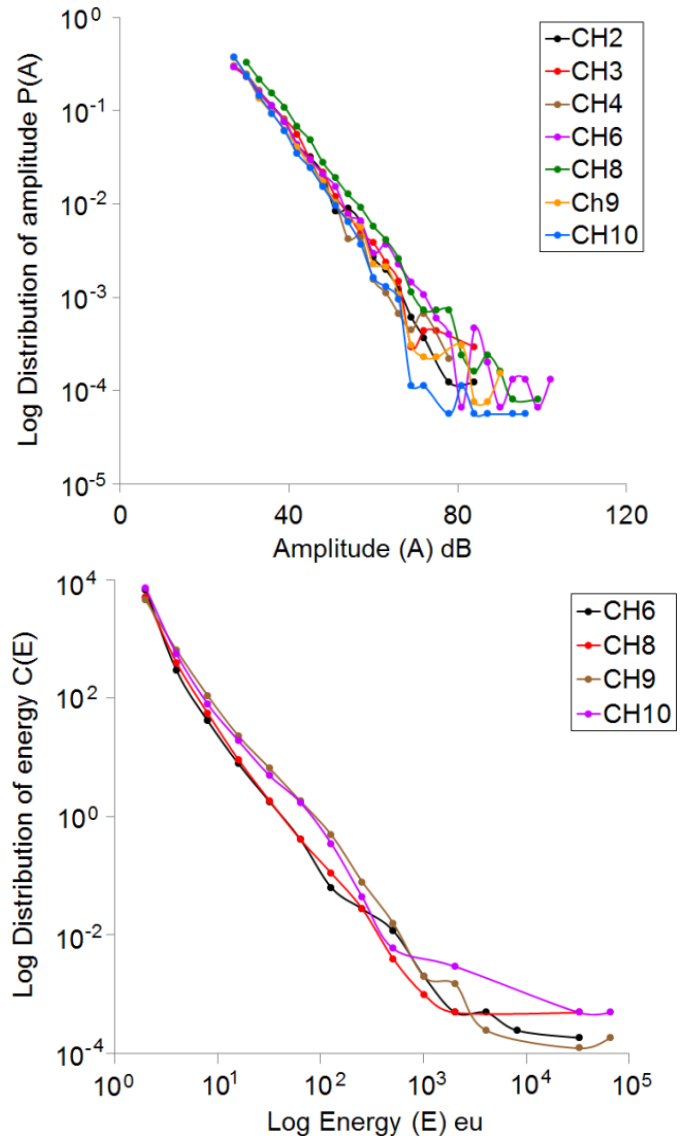

Fig. 3. AE amplitude (A) distribution and energy (E) distribution during the fracturing test on Saltwash North Sandstone sample.

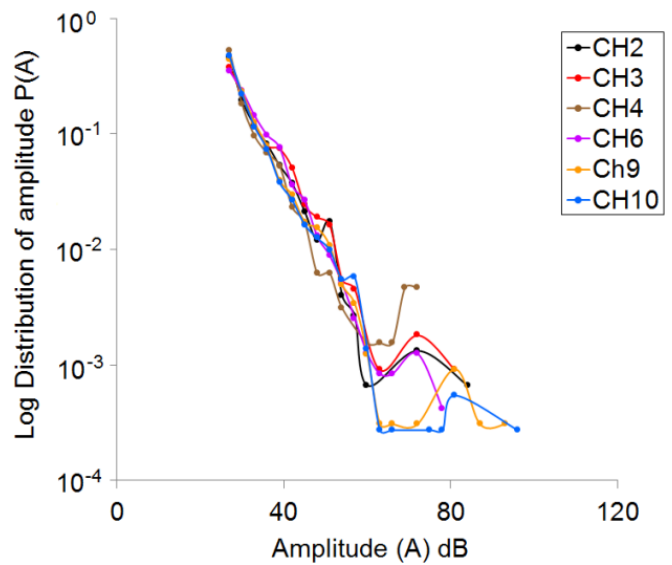

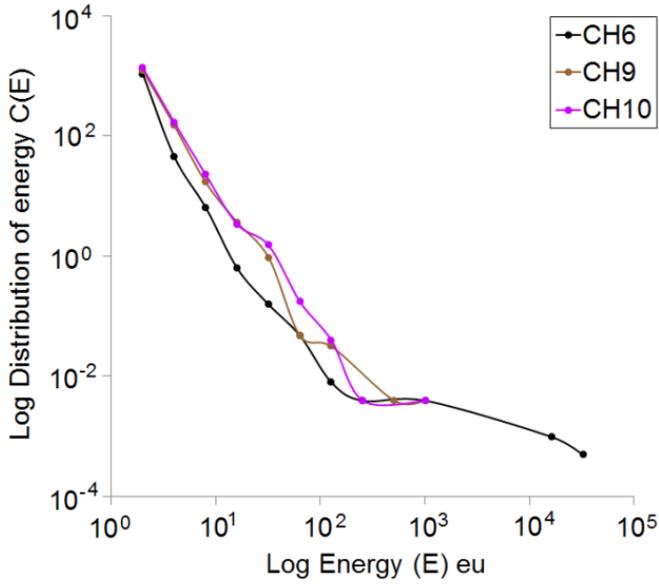

Fig. 4. AE amplitude (A) distribution and energy (E) distribution during the fracturing test on Lixhe Chalk sample.

Sandstone and one for Chalk) of the statistical distributions are shown in Fig. 3 and Fig. 4. It seems that the AE amplitudes follow an exponential distribution

$$P(A) \sim e^{-A/\alpha}, \qquad (1)$$

and the AE energies follow a power law distribution

$$C(E) \sim E^{-\beta}, \qquad (2)$$

for all the rock types. But the values of $\alpha$ and $\beta$ differ from rock type to rock type. We present these exponents values in a table below:

Tab. 1. Distribution exponents for AE amplitude and Energies for different rock types

| Rock type | $\alpha$ | $\beta$ |
|---|---|---|
| Barea | 6 | 1.7 |
| Castlegate | 7 | 1.7 |
| Red Wildmoor | 6 | 1.6 |
| Saltwash North | 7 | 1.8 |
| Saltwash South | 6 | 1.8 |
| Mons chalk | 6 | 1.4 |
| Lixhe chalk | 5 | 1.9 |

High $\beta$ values for chalk indicate that high energy events are less populated in chalk samples – which is consistent with the fact that weak rocks do not produce big acoustic bursts. One can verify this fact by comparing Fig. 3 and Fig. 4.

## 4. TIME EVOLUTION OF RADIAL STRAIN

The hold period at the initial part of the tests reveal a significant amount of creep, which may disturb the interpretation of the strain data. Creep can be evaluated using a model that combines a spring and dashpots elements (modified Burgers substance) [1]. This model takes into account transient creep and steady state creep. According to this model, creep during loading can be represented mathematically as:

$$\varepsilon = a \cdot (1 - e^{-t/\tau}) + b \cdot t, \qquad (3)$$

where $a$ is the amplitude and $\tau$ is the time constant of the transient creep, and $b$ is the steady state creep velocity. Fig. 5 (a) shows as an example how the model matches with the observations, while Fig. 5 (b) shows how the radial strain develops when the delayed deformation is subtracted in accordance with this model. The creep corrected data gives a better description of the immediate response to borehole pressure changes. One can notice that the corrected radial strain vs. time plot shows a significance change of its slope around the fracturing point and the rate of AE events increases rapidly in that area (compare Fig. 2 and Fig. 5). Creep estimation parameters obtained for all rock types are given in Tab. 2.

Tab. 2. Parameter describing the time evolution of the radial creep strain

| Rock type | $a$ | $\tau$ | $b$ |
|---|---|---|---|
| Barea | 2.5e-2 | 62.5 | 4.9e-5 |
|  | 1.5e-2 | 30.6 | 4.4e-5 |
| Castlegate | 4.0e-2 | 71.3 | 4.7e-5 |
|  | 2.0e-2 | 20.4 | 2.2e-4 |
| Red Wildmoor | 2.8e-2 | 44.4 | 3.5e-5 |
|  | 3.0e-2 | 46.8 | 1.1e-4 |
|  | 3.6e-2 | 54.1 | 7.2e-5 |
| Saltwash North | 5.4e-2 | 72.1 | 7.9e-5 |
|  | 4.4e-2 | 42.5 | 5.8e-5 |
| Saltwash South | Large data scattering | | |
|  | Large data scattering | | |
| Mons chalk | 4.4e-3 | 12.1 | 4.2e-5 |
|  | 9.8e-3 | 12.3 | 6.6e-5 |
| Lixhe chalk | 1.2e-2 | 23.4 | 2.1e-5 |
|  | 2.1e-2 | 53.7 | 3.5e-5 |
|  | 1.6e-2 | 22.9 | 1.4e-4 |

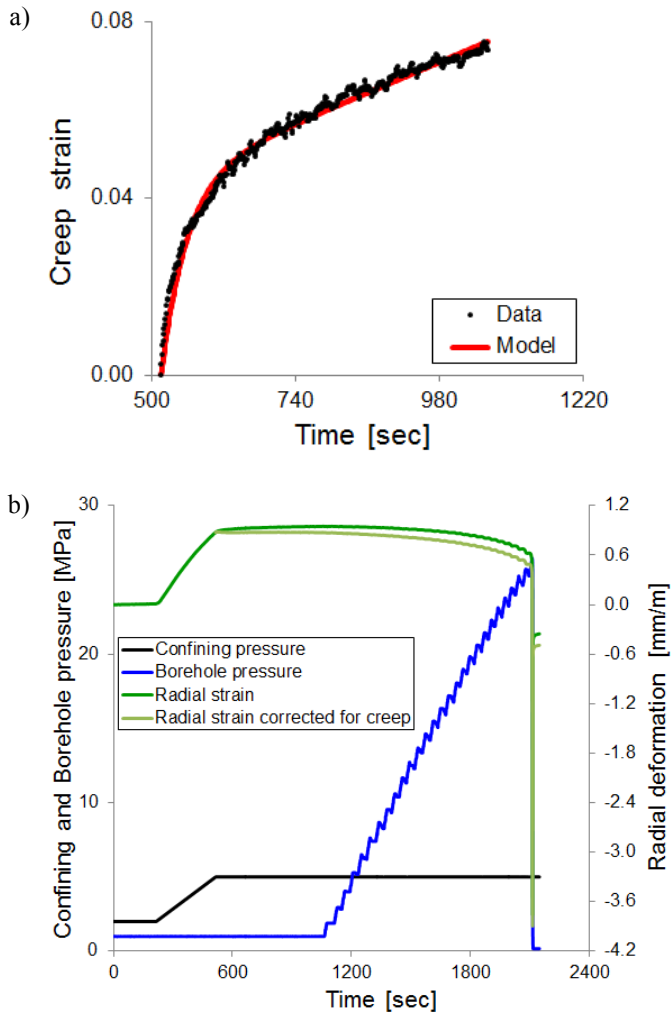

Fig. 5. Time-evolution of radial creep during a fracturing test (a) and relevant radial strain correction for creep (b). The data refers to a test on Saltwash North sandstone.

## 5. DISCUSSIONS AND CONCLUSIONS

We have studied fracturing behavior of 7 rock samples through laboratory tests and post-test AE and image analysis. Statistical analysis of AE events gives distribution exponents for AE amplitude and AE energies – these exponents differ from rock type to rock type. High energy distribution exponent ($\beta$ values) for weak rocks (chalks) is a signature that high energy events are less populated in weak samples – which is consistent with the observation that weak rocks do not produce big acoustic bursts. After subtracting the creep part, the corrected radial strain vs. time plot gives the actual response of the sample against increased borehole pressure. The slope of the plot changes rapidly around the fracturing point – which indicates significant damage of the rock sample before complete fracturing. We have started analysing this fracturing scenario through discrete element modelling (DEM) putting some exact input parameters like tensile strength of the rocks, borehole pressure, element breaking criteria etc. The model results match well with that of the lab- tests qualitatively. We are now going to calibrate the radial strain vs. time plot produced in DEM code – against the same from lab test for different types of rocks. The aim of this study is to find out the actual scaling factor (sample size dependent) that can give us exact calibration of the plot – from which we can estimate the fracture length vs. radial deformation (or borehole pressure) for different rock types.

**Acknowledgements:** This work is supported by funding from Research Council of Norway (NFR) through grant no. 199970/S60 and 217413/E20.